
\bigskip

\rightline{\bf ILG-TMP-93-02}

\rightline{\bf April, 1993}

\bigskip

\bigskip

\bigskip

\centerline{{\bf Alexander A.Belov}\footnote{$^{\heartsuit}$}{e-mail:
mitpan@adonis.ias.msk.su [Belov]}}

\smallskip

\centerline{Int.Inst.for Math.Geophys.and Predictions Theory}

\centerline{Warshavskoe sh. 79, k.2 Moscow 113556 Russia}

\bigskip

\bigskip

\centerline{{\bf Karen D.Chaltikian}
\footnote{$^{\diamondsuit}$}{e-mail: mitpan@adonis.ias.msk.su [Chaltikian],
chalt@cpd.landau.free.msk.su}}

\smallskip

\centerline{Landau Institute for Theoretical Physics}

\centerline{Vorobyevskoe sh.2, Moscow 117940 Russia}

\bigskip

\bigskip

\centerline{\bf LATTICE\quad ANALOGUE\quad OF\quad W-INFINITY\quad ALGEBRA}

\bigskip

\centerline{\bf AND\quad DISCRETE\quad KP-HIERARCHY}

\bigskip

\bigskip

\centerline{\bf Abstract}

\bigskip
In development of the started activity on lattice analogues of $W$-algebras,
we define the notion of lattice $W_{\infty}$-algebra, accociated with lattice
integrable system with infinite set of fields. Various kinds of reduction to
lattice $W_N$-algebras, related to discrete $N$-KdV hierarchies are described.
We also discuss the connection of our results with those obtained in the

papers of Xiong [13] and Bonora [14].

\def\H#1{{\cal H}^{(#1)}}
\def\h{\cal H}

\centerline{\bf 0. Introduction}

\bigskip

     Recently there has been a  great  interest  in  the  lattice
analogues of 2D field  theory models  possessing  conformal
invariance [1-4]. For the first  time  this  interest  arised  in
connection with Liouville model [1-3].  As  is  well  known,  its
quantization in continuous limit is rather difficult in the range
of central charges $1<c<25$, where  usual  perturbation  theory  is
inapplicable. 'Latticization' of the world sheet can be viewed as
some   alternative   to   usual   point-splitting    method    of
regularization.  That's  why  lattice  conformal  theories   give
another way of quantization, more convenient  in  some  cases  in
comparison with  that of based on normal  ordering  procedure.
An additional motivation for studying  the  lattice  analogue  of
conformal invariance comes from the traditional understanding  of
Integrable  Massive  models  as   appropriate   deformations   of
Conformal  ones  [5-7].  It  seems  highly  desirable  that  this
important  concept  be  formulated    on   purely   lattice
language.

In the previous paper [8] authors presented a natural lattice analogue
 of $W_3$-algebra.  This algebra has been obtained as $W_3$-generalization
of the Faddeev-Takhtadjian-Volkov (FTV) algebra [1], which is considered
presently as lattice analogue of Virasoro algebra. In the paper [8]
we proposed some general method to obtain classical and quantum
versions of lattice $W$-algebras ($LW$), based on Feigin construction
of lattice screening charges [9]. This method in principle guarantees
the possibility to obtain any $LW$-algebra with finite number of fields
with integer spins. Besides, considering such an algebra as {\it second}
Poisson structure one can easily find the corresponding{\it first} structure
 and a family of involutive integrals of motion. Further on {\it true} lattice
analogues of $W_N$ algebras (second structure) will be referred to as $LW_N$
and the corresponding first structure to as $L_{*}W_N$.
The main drawback of the abovementioned construction is that it does not
look to admit straightforward
generalization for the case of the infinite number of fields. The aim
 of present paper is to declare an alternative method of building
$L_{*}W_N$-algebras, based on certain reductions of some {\it guessed}
\footnote{$^1$}{An algebra related to ours via some {\it involutive}
transformation recently has been  obtained in
the paper of Xiong [13] . Detailed
discussion is given in Sect.3} lattice analogue of the {\it first}
 Poisson structure for $W_{\infty}$ and to find its relation with
some integrable models.

The detailed plan of the paper is as follows.
In Sect.1 we remind the main results of the paper [8], concerning
free field representation of $LW_3$ algebra, and its bi-hamiltonian
structure which led us to the  the notion of $L_{*}W_{\infty}$ and give some
additional motivations of the work.  In Sect.2 we discuss some
interesting properties of the Toda Lattice Hierarchy (TLH), which
turn out to be closely related with the notion of "lattice conformal
symmetry" and certain $L_{*}W_{\infty}$ reductions. In
Sect.3 we give the central definition of $L_{*}W_{\infty}$ and $LW_{\infty}$,
 discuss two various kinds of their reductions to the algebras with finite
number of fields.  We also compare our results with those of
Xiong [13] and Bonora [14].
Sect.4 is devoted to (bi-)hamiltonian structure of the integrable
model, possessing $LW_{\infty}$-symmetry and Lax representaion of the
model and discuss its connection with
both types of reduction. In the Appendix we bring some useful  formulas
concerning the multihamiltonian structure of TLH.

\bigskip
\centerline{\bf Acknowledgments}

\medskip
 Authors are greatly indebted to Prof. I.M.Krichever
 for valuable discussions and very useful remarks. We also thank
Prof. A.S.Mezhlumian for the help in preparation of the text
for publication.

\bigskip

\bigskip

\bigskip

\centerline{\bf 1. $LW_3$-algebra -- starting point.}

\bigskip

This section contains some basic definitions and results and can be considered
as practical introduction in the subject. We start from
Feigin construction of lattice screening vertex operators. Basic lattice
variables being the direct analogues of continuum screening charges [10] are
defined via their commutation relations as follows:
$$a_i(m)a_j(n)=q^{A_{ij}}a_j(n)a_i(m),\hbox{\ when\ }n>m\eqno(2^{\prime})$$
\centerline{and}
$$a_i(n)a_{i+1}(n)=q^{-1}a_{i+1}(n)a_i(n)$$
$$a_i(n)a_j(n)=a_j(n)a_i(n),\hbox{\ when\ }\mid i-j\mid\geq 2$$
where $A_{ij}$ is Cartan matrix of the appropriate semisimple Lie algebra.
Note that $q$ here can be treated as free  parameter  contrary  to the
continuous case where it was rigidly determined. Correspondingly,
lattice analogues of the screening charges are defined as
$$Q_i=\sum_na_i(n)$$
 For the case of $SL_3$
 we denote basic lattice vertices by $b_n$ and $c_n$. Further on we will
discuss only classical case, the original Poisson brackets (PB) becoming
$$\{b_n,b_m\}=\epsilon (m-n) b_nb_m \qquad \qquad
\{c_n,c_m\}=\epsilon (m-n) c_nc_m$$
$$\{b_n,c_m\}=\theta (n-m) b_nc_m$$
where $\epsilon (n)$ and $\theta (n)$ are standardly defined as
$$\epsilon (n)=\cases{1, &if $n>0$\cr 0, &if $n=0$\cr -1, &if $n<0$}\qquad
 \theta (n)=\cases{1, &if $n\ge 0$\cr 0, &otherwise}$$
Going further in parallel with the continuum limit, Feigin proposed to define
$Vir$ and $W$ lattice generators as 3- and 4-point objects lying in the
{\bf zero-graded part of  kernel of the adjoint action of lattice
screening charges}, the gradation being defined as ($\bar a\equiv a^{-1}$):
        $$ deg(a_r(n))=1 \qquad \qquad     deg({\bar a}_r(n))=-1$$

Introducing more convenient variables $p_n$ and $d_n$
$$p_n \equiv {\bar b}_n b_{n+1}\qquad \qquad \hbox{and}\qquad \qquad d_n \equiv
c_n{\bar c}_{n+1}$$
with the local PB's
$$\{ p_n,p_{n+1}\}=p_np_{n+1} \qquad \qquad \{ p_n,d_n\}=p_nd_n$$
$$\{ d_n,d_{n+1}\}=d_nd_{n+1} \qquad \qquad \{ p_{n+1},d_n\}=-p_{n+1}d_n
\eqno(1)$$
after some calculations we obtain the following expressions for bi- and
three-local (in terms of original vertices) generators [8]:
$$L_n=(p_{n+1}d_n+p_{n+1}+d_n)(1+p_n+d_n)^{-1}(1+p_{n+1}+d_{n+1})^{-1}$$
$$W_n=d_np_{n+2}(1+p_n+d_n)^{-1}(1+p_{n+1}+d_{n+1})^{-1}
(1+p_{n+2}+d_{n+2})^{-1}
\eqno(2)$$
These fields form closed algebra $LW_3$ [8]:
$$\{L_n,L_{n+1}\}=(L_nL_{n+1}-W_n)(1-L_n-L_{n+1})\qquad \qquad
\{L_n,L_{n+2}\}=-L_nL_{n+1}L_{n+2}+W_nL_{n+2}+W_{n+1}L_n$$
$$\{L_n,W_n\}=-W_nL_nL_{n+1}+W_n^2=\{W_n,L_{n+1}\}$$
$$\{W_n,L_{n+3}\}=-W_nL_{n+2}L_{n+3}+W_nW_{n+2}\qquad \qquad
\{W_n,L_{n+2}\}=W_nL_{n+2}(1-L_{n+1}-L_{n+2})+W_nW_{n+1}$$
$$\{L_n,W_{n+1}\}=L_nW_{n+1}(1-L_n-L_{n+1})+W_nW_{n+1}\qquad \qquad
\{L_n,W_{n+2}\}=-L_nL_{n+1}W_{n+2}+W_nW_{n+2}$$
$$\{W_n,W_{n+1}\}=W_nW_{n+1}(1-L_n-L_{n+2})\qquad \qquad
\{W_n,W_{n+2}\}=W_nW_{n+2}(1-L_{n+1}-L_{n+2})$$
$$\{W_n,W_{n+3}\}=-W_nW_{n+3}L_{n+2}\eqno(3)$$
In continuous limit, setting $x\equiv n\Delta$
$$W_n\;\longrightarrow \;{1 \over 27}
(1-{\Delta}^2u(x)-{{\Delta}^3\over 2}w(x))$$
$$L_n\;\longrightarrow \;{1\over 3}(1-{{\Delta}^2\over 3}u(x))$$
one can easily check that the fields $u(x)$ and $w(x)$
form classical $W_3$-algebra.

Indeed, the same path beginning from the single vertex $a_n$, can be passed in
order to obtain FTV algebra [9], appeared first in the paper on lattice
Liouville theory and Volterra model [1]:
$$\{S_n,S_{n+1}\}=S_nS_{n+1}(1-S_n-S_{n+1})$$
$$\{S_n,S_{n+2}\}=-S_nS_{n+1}S_{n+2} \eqno(4)$$
where $S_n$ has the following representation ($p_n={\bar a}_na_{n+1}$) [1,2,9]:
$$S_n=p_{n+1}(1+p_n)^{-1}(1+p_{n+1})^{-1}$$
Strange as it might seem that the fields $L_n$ do not form a subalgebra in
$LW_3$.
However, looking attentively at the first line of commutation relations (3),
one can see that setting $W_n=0$ one obtains FTV algebra (4).

The model with symmetry (3), as well as Volterra model [3], turns out to
 possess multihamiltonian structure [8]. Namely, one can set the following
 first structure on the space of the fields $L_n$ and $W_n$, viewing (3)
as the {\it second} one:
$${\{L_n,L_{n+1}\}}_1=L_nL_{n+1}-W_n \qquad \qquad
{\{W_n,W_{n+1}\}}_1=W_nW_{n+1}$$
$${\{L_{n+2},W_n\}}_1=-L_{n+2}W_n\qquad \qquad
{\{L_n,W_{n+1}\}}_1=L_nW_{n+1}$$
$${\{W_n,W_{n+2}\}}_1=W_nW_{n+2}\eqno(5)$$
This algebra will be further referred to as $L_{*}W_3$.

One can easily check, that the following are the involutive hamiltonians of
the model
$${\cal H}^{(1)}=\sum_n L_n$$
$${\cal H}^{(2)}=\sum_n({L_n^2 \over 2}+L_nL_{n+1}-W_n)$$
$${\cal H}^{(3)}=\sum_n \left( {L_n^3 \over 3}+L_n^2L_{n+1}+L_nL_{n+1}^2+
L_nL_{n+1}L_{n+2}-W_n(L_{n+1}+L_{n+2})-L_n(W_n+W_{n+1})\right)$$
related by the recursion
$${\{{\cal H}^{(i)},\Psi \}}_2={\{{\cal H}^{(i+1)},\Psi \}}_1$$
where $\Psi$ is any field from the set $\{L_n, W_n\}$.

\bigskip

\bigskip

\bigskip
\centerline{\bf 2. Doubling transformation in reduced TLH.}

\bigskip

In this section we briefly discuss some interesting properties of
the well-known TL system [11,12] which turn out to be relevant for the subject
involved. We start from the  Hamiltonian
$${\h}_T=\sum_n(U_n+{p_n^2\over 2})\eqno(6)$$
where $U_n=e^{q_n-q_{n-1}}$ and  Poisson structure is standard
$$\{p_n,q_m\}=\delta_{n,m}\eqno(7)$$
This system possess multihamiltonian structure [11-13], and one can easily
obtain the recursion operator and higher Poisson structures (complete
 derivations are brought in the Appendix) . It will be convenient for us
to refer to the original PB (7) as ${1\over 2}$-bracket. Then the {\it first}
bracket of TL is given by
$$\{U_n,U_{n+1}\}=-U_nU_{n+1}\qquad \qquad \{p_n,p_{n+1}\}=-U_{n+1}$$
$$\{U_n,p_n\}=-U_np_n\qquad \qquad \{U_{n+1},p_n\}=U_{n+1}p_n\eqno(8)$$

Our further interest will be in the "restricted" model with $p_n=0$, or
in other words, with all "{\it half-integer}" times equal to zero. In this
case, as is well-known TL becomes the Volterra model [12]. Through the
recursion procedure (A.4) one can find its {\it second} structure
to be the FTV algebra, which we write here in a more convenient than
 (4) but equivalent form
$${\{U_n,U_{n+1}\}}_2=-U_nU_{n+1}(U_n+U_{n+1})$$
$${\{U_n,U_{n+2}\}}_2=-U_nU_{n+1}U_{n+2}\eqno(9)$$
We note that the canonical FTV algebra can be obtained either as a linear
combination of the compatible structures (9) and (8) or as a {\it second}
structure for the system of hamiltonians, differing from the original ones by
some linear transformation.

The recursion procedure provides us also with ${3\over 2}$-bracket
(see Appendix) but we do not bring it here as it is identical zero
on the surface $p_n=0$ [13].

Now we describe the {\bf doubling} transformation of the TLH.
For this introduce the notations
$$u_n=U_{2n} \qquad \qquad v_n=U_{2n+1}\eqno(10)$$
and define the fields
$$p_n^{(1)}\equiv u_n+v_n \qquad \qquad U_n^{(1)}\equiv v_{n-1}u_n$$
One readily checks that these fields form algebra (8). Indeed, this
 "doubling" procedure can be carried on again. One can obtain TLH's for
the set of pairs $(p_n^{(i)}, U_n^{(i)})$. Thus, a separate interesting
problem of studying this system as a whole arises. However, as this is not
the subject of the present paper, we will not elaborate on it here.

Let us consider another realization of the algebra (8) from the "doubled"
variables, introducing the generators as
$$l_n\equiv p_n \qquad \qquad w_n\equiv u_nu_{n+1}+u_nv_{n+1}+v_nv_{n+1}$$
The relations have the form
$$\{l_n,l_{n+1}\}=-l_nl_{n+1}+w_n \qquad \qquad \{l_n,w_n\}=0$$
$$\{l_n,w_{n+1}\}=l_{n+2}(w_n-l_nl_{n+1})\qquad \qquad
\{l_{n+2},w_n\}=l_n(l_{n+1}l_{n+2}-w_{n+1})$$
$$\{w_n,w_{n+1}\}=-(l_nl_{n+1}-w_n)(l_{n+1}l_{n+2}-w_{n+1})$$
$$\{w_n,w_{n+2}\}=l_nl_{n+3}(w_{n+1}-l_{n+1}l_{n+2})\eqno(11)$$
This algebra will be obtained in next section from the certain reduction of
$L_{*}W_{\infty}$.

\bigskip

\bigskip

\bigskip
\centerline{\bf 3. Universal algebras $L_{*}W_{\infty}$ and $LW_{\infty}$.}

\bigskip

In this section we give the universal generalization of the first structures
(5) and (8) to the case of infinite number of fields with spins running from
 2 to $\infty$. This algebra have been called $L_{*}W_{\infty}$
in previous sections.
The objects of study in this section will be the {\bf integer spin fields}
(ISF) ${\{A_n^{(p)}\}}_{p=1}^{\infty}$ with spins $p+1$.  We also use
formal notations $A_n^{(0)}\equiv 1$ and $A_n^{(p<0)}\equiv 0$.
In examples we will write
$$A_n^{(1)}\equiv A_n, \qquad \qquad A_n^{(2)}\equiv B_n, \qquad \qquad
A_n^{(3)}\equiv C_n,$$
$$A_n^{(4)}\equiv D_n, \qquad \qquad A_n^{(5)}\equiv E_n,\qquad \ldots$$

The name ISF is inspired by the
continuous limit in which the  fields $A_n^{(p)}$ have the following
form
$$\eqalign{&A_n\to 1\oplus w_2\Delta^2\oplus O(\Delta^3)\cr
&B_n\to 1\oplus w_2\Delta^2\oplus w_3\Delta^3 \oplus O(\Delta^4)\cr
&C_n\to 1\oplus w_2\Delta^2\oplus (w_3\oplus w_2')\Delta^3
\oplus w_4\Delta^4\oplus O(\Delta^5)\cr
&D_n\to 1\oplus w_2\Delta^2\oplus (w_3\oplus w_2')\Delta^3 \oplus
(w_4\oplus {w_3}'\oplus {w_2}''\oplus w_2^2)\Delta^4\oplus w_5\Delta^5
\oplus O(\Delta^6)\cr
&\ldots \qquad \ldots \qquad \ldots\cr}$$
where $O(\Delta^m)$ means the terms irrelevant for the purpose
of obtaining the continuos limit and $w_k(x)$ are the classical analogues
of quantum fields $W^{(p)}(x)$ forming $W_{\infty}$-algebra, $W^{(2)}(x)
\equiv T(x)$ being energy-momentum tensor and having the OPE of the form
$$T(x)\;W^{(p)}(z)={pW^{(p)}(z)\over {(x-z)}^2}+
{\partial W^{(p)}(z)\over x-z}$$

\bigskip

\bigskip
\leftline{\bf3.a. First bracket for the ISF.}

\medskip
We propose the following natural generalization of the {\it first} Poisson
structures (5) and (8) for the case of infinite number of fields.
$$\{A_n^{(p)},A_{n+m}^{(q)}\}=\theta_m^{p,q}\cdot
(-A_n^{(p)}A_{n+m}^{(q)}+A_n^{(q+m)}A_{n+m}^{(p-m)})\eqno(12)$$
where $\theta$-function is defined as a sum of $\delta$-functions over the
appropriate interval
$$\theta_m^{p,q}\equiv \theta (0\leq p-m \leq q-1)\equiv
\sum_{k=0}^{q-1}\delta_{k,p-m}$$
Proof of Jacobi identity is straightforward and relies on the three main
properties of this $\theta$-function:
$$\eqalign{&\theta_m^{p,q}=\theta_{m+k}^{p+k,q}\cr
&\theta_{m-k}^{p,q+k}=\theta_m^{p,q}+\theta_{p+1}^{m,k}\cr
&\theta_{m+k}^{p,q-k}\delta_{q,k+r}=\theta_m^{p,q}\;\theta_k^{p-m,r}-
\theta_k^{q,r}\;\theta_{m+k}^{p,q-k}-
\theta_{m+k}^{p,r}\;\theta_m^{r+m+k,q}\cr}$$
We would like to note also that Jacobi identity for this algebra has
the form of
classical Yang-Baxter equation with discrete spectral parameter. To see this it
is sufficient to rewrite (12) in the form
$$\{A_n^{(p)},A_{m}^{(q)}\}=r_{r\,s}^{p\,q}(m-n)A_n^{(r)}A_m^{(s)}\eqno(13)$$

Now we are in a position to define the first type of reduction, which we call
 $L_{*}W_N$. Namely, we simply set the constraints
$$A_n^{(p)}=0, \qquad \hbox{for all} \qquad p\geq N\eqno(14)$$
The obtained algebra $L_{*}W_N$ defines the first Poisson structure for
the lattice analogue of the so-called $N$-th KdV hierarchy. In particular,
algebras (5)--$L_{*}W_3$ ($A_n\equiv L_n,\;B_n\equiv W_n$)
and (8) (for $p_n=0$)--$L_{*}Vir$ can be obtained
in this way. Surely, knowing the algebra $L_{*}W_N$ and full set of its
Hamiltonians (see next section for explicit expressions), one can easily
obtain {\it the true} $LW_N$-algebra, like FTV ($=LVir$) and $LW_3$, which
in continuous limit becomes  classical $w_N$-algebra, corresponding
to  $N$-KdV integrable sytem. Clearly, this algebra gives the {\it second}
bracket for  $N$-KdV lattice hierarchy. General formula also can be obtained
and has the form:
$$\eqalign{{\{A_n^{(p)},A_{n+m}^{(q)}\}}_2=&
-A_n^{(p)}A_{n+m-1}^{(1)}A_{n+m}^{(q)}-A_n^{(p)}A_{n+p}^{(1)}A_{n+m}^{(q)}\cr
&+A_n^{(p)}A_{n+m-1}^{(q+1)}+A_n^{(p+1)}A_{n+m}^{(q)}\cr
&+A_n^{(q+m)}A_{n+m-1}^{(1)}A_{n+m}^{(p-m)}+
A_n^{(q+m)}A_{n+p}^{(1)}A_{n+m}^{(p-m)}\cr
&-A_n^{(q+m)}A_{n+m-1}^{(p-m+1)}-A_n^{(q+m)}A_{n+m}^{(p-m+1)}\cr}\eqno(15)$$
but there are some technical difficulties for several particular commutators
of the operators sitting on the adjacent sites. These commutators (i.e. with
$m=0,\pm 1,p,p\pm 1$) are given by slightly different formulas. We do not bring
them here for the lack of place;
however, for any finite $N$ all the 2-brackets can be easily computed from the
bihamiltonian recursion relation, as all the integrals of motion are known for
the system (see Sect. 4).

Ending this paragraph, it is worth noting, that
for any $LW_N$-algebra with finite $N$ one might expect that {\it all} the
fields $A_n^{(p)}$ should be expanded to $N$-th order in lattice constant;
however,
as we have established earlier, it is sufficient to consider only
{\it first $p$ terms} in the expansion of $A^{(p)}$ in order to obtain the
desired object $LW_N$.

\bigskip

\bigskip
\leftline{\bf 3.b. Second type of reduction and Xiong-Bonora algebra [13,14]}

\medskip

First of all define the transformation from ISF to the {\bf string
generating fields} (SGF) as follows:
$$\eqalign{&L_{n,n}=A_n^{(1)}\cr
&L_{n,n+1}=A_n^{(1)}A_{n+1}^{(1)}-A_n^{(2)}\cr
&L_{n,n+2}=A_n^{(1)}A_{n+1}^{(1)}A_{n+2}^{(1)}-A_n^{(1)}A_{n+1}^{(2)}-
A_n^{(2)}A_{n+2}^{(1)}+A_n^{(3)}\cr
&\cdots \qquad \cdots \qquad \cdots \cr
&L_{n,n+p-1}=\sum_{\vec{m}\in \cal P} \prod_{j=0}^{|\vec{m}|-1}
A_{n+M_j}^{(m_{j+1})}\;{(-1)}^{p+|\vec{m}|}\cr}\eqno(16)$$
where $\cal P$ is the set of partitions of segment $[n,n+p]\cap Z$ and other
notations are
$$\vec{m}=(m_1,m_2,\ldots,m_s), \qquad s=|\vec{m}|,$$
$$ \sum_{l=1}^sm_l=p, \qquad M_j=\sum_{l=1}^jm_l,\qquad M_0=0$$
The main  interesting property of this transformation is that is
 {\it involutive}, i.e. if we denote the r.h.s. of eq.(16) by
${\cal F}_p(\{A_m^{(q)}\})$ than one has
$$A_n^{(p)}={\cal F}_p(\{L_{m,m+q}\})\eqno(17)$$

The original first structure (12) induces some 1-bracket on the SGF's
and direct calculation shows that bracket to be nothing but Xiong-Bonora
algebra (see formula (15) of Ref.13 or formula (3.40) or ref.14)
if one identifies our SGF's with the fields from Ref.14 as
$$L_{n,n+p}\equiv a_{p+1}(n)$$

Now the reduction which we call $L^{*}W_N$ ($W$ now only reminds us about
its origination from $L_{*}W_{\infty}$) is defined by setting the
{\it constraints} on the ISF,
which can be very simply written in terms of SGF's (16)
$$L_{n,n+p}=0 \qquad \hbox{for all} \qquad p\ge N\eqno(18)$$
This reduction for $N=2$ gives us an abelian algebra
$$\{A_n,A_{m}\}=0$$
In the case of $N=3$ we obtain the algebra (11)
($A_n\equiv l_n,\;B_n\equiv w_n$)
 about which we have known from previous section that it gives TLH 1-bracket
written in "doubled" space (10), if one rewrite it in terms of SGF's
substituting the expressions (17) instead of $A_n^{(p)}$. One should
mention here that the name "string generating field" comes from the
explicit realization of the $N=3$ case in
terms of the "doubled" space variables (10), in which the original ISF's
are expressed via the recursion relation
$$\eqalign{A_n^{(p+1)}&=u_nA_n^{(p)}+v_nv_{n+1}\ldots v_{n+p}\cr
&=u_nu_{n+1}\ldots u_{n+p}+A_n^{(p)}v_{n+p}\cr}\eqno(19)$$

Operating with the SGF's from the
very beginning Xiong [13] obtained TLH $1$- and ${3\over 2}$-brackets and
showed
that some appropriate combination of those gives two commuting Virasoro
algebras
in the continuous limit. However, his scheme failed in obtaining true
{\it second}
Poisson structures for $N>3$ ($LW_N$-algebras), and so we believe that an
appropriate language here is that of ISF's, where one obtains true first
brackets for lattice $N$-KdV hierarchies. We believe, in particular, that
true lattice Virasoro
is an algebra closed on one field, i.e. FTV algebra which dtermines the
 {\it second}
Poisson structure of the TLH {\it on the surface} $p_n=0$, rather than
the abovementioned linear combination of its $1$- and ${3\over 2}$-structures.

\bigskip

\bigskip

\bigskip
\centerline{\bf 4. Hamiltonian structures and Lax representation.}

\bigskip

In this section we describe  the Lax representation of the integrable system
associated with $LW_{\infty}$. Formula for $L$-operator also has been obtained
in [13]. Its upper triangular matrix elements are given by the set of
SGF's $L_{n,n+p}$ (16), and the lower ones are defined as
$$L_{n,n-k}={\delta}_{k,1}, \qquad k>0$$
 Hamiltonians as usual are given by
$$\H{k}={1\over k}Tr(L^k)$$
and coincide with those of $N$-KdV hierarchy under the constraints (14).

Again we emphasize the advantage of the ISF's language. In Ref.14 it has been
noticed, that equations of motion for $L_{n,n+p}$ in the $N=4$ case on the
surface $L_{n,n}=0$ cannot be identified with discrete Boussinesq hiearchy as
is naturally expected from the multimatrix models intuition
because of wrong continuous limit. We see now, that it is not too surprisingly,
because only ISF fields give true realization of the lattice $N$-KdV hierarchy.
For example,  for $N=3$
Hamiltonians coincide with those of lattice Boussinesq hierachy (see Sect.1)
and, correspondingly, equations of motion have true continuous limit [8].

\bigskip

\bigskip

\bigskip
\centerline{\bf 5. Conclusion}

\bigskip

In conclusion, we have built the integrable system with infinite number
of fields. This system possesses first and second  compatible Poisson
structures, the latter being the lattice analogue of $W_{\infty}$ algebra.
 Certain reduction of this system to $N<\infty$ of fields gives us
first and second Poisson structures of lattice $N$-KdV hierarchies. Phase
space of the model in this case can be parametrized by $N$ lattice
currents as in the example of Sect. 1. There also exists another reduction,
in which, generally all the fields have non-trivial, although as we have
seen in  Sect.3.b not independent evolutions. The reduction here is revealed
in that the phase space of the model is parametrized by only few variables.
A good example is $N=3$ case, in which we obtain Toda lattice hierarchy
after the substitution (19). It is noteworthy, that in
this particular case we deal with the phenomena analogous to that of
the so-called two-boson realization of KP hierarchy in continuous limit [15].

\bigskip

\bigskip

\bigskip
\centerline{\bf Appendix. Multihamiltonian structure of the TLH.}

\bigskip

Below some useful instrumentary concerning TLH is presented. First we
introduce the following block-matrix form of $j$-th symplectic structure:
$$\Omega_{mn}^{(j)}\equiv \pmatrix{{\{p_m,p_n\}}_j&{\{p_m,u_n\}}_j\cr
{\{u_m,p_n\}}_j&{\{u_m,u_n\}}_j\cr}\eqno(A.1)$$
Original symplectic structure $\Omega^{({1\over 2})}$ (eq.(7)) in this
notation is given by
$$\Omega_{mn}^{({1\over 2})}=\pmatrix{0&u_n(\delta_{n,m}-\delta_{n,m+1})\cr
-u_m(\delta_{m,n}-\delta_{m,n+1})&0\cr}\eqno(A.2)$$
{\it First} Poisson structure (eq.(8)) is
$$\Omega_{mn}^{(1)}=\pmatrix{u_m\delta_{m,n+1}-u_n\delta_{m,n-1}&
p_mu_n(\delta_{m,n}-\delta_{m,n-1})\cr p_nu_m(\delta_{m,n+1}-\delta_{m,n})&
-u_mu_n(\delta_{m,n-1}-\delta_{m,n+1})\cr}\eqno(A.3)$$

One can give the following definition of the {\bf Recursion Operator}
$${\cal R}\;\equiv\; \Omega^{(1)} \cdot {\Omega^{({1\over 2})}}^{-1}$$
with explicit form
$${\cal R}=\pmatrix{p_n\delta_{k,n}&\delta_{k,n}+{u_{n+1}\theta
 (n-k+2)-u_n\theta (n-k+1)\over u_k}\cr
 u_n(\delta_{k+1,n}+\delta_{k,n})&p_n\delta_{k,n}+{(p_n-p_{n-1})
 u_n\theta (n-k)\over u_k}\cr}$$
By construction the following relation is hold:
$${\vec{v}}^{(j)}\cdot {\cal R} ={\vec{v}}^{(j+{1\over 2})}$$
where ${\vec{v}}^{(j)}$ stands for vector field generated by $j$-th
Hamiltonian $v^{(j)}_n=({\partial \H{j}\over \partial p_n},
{\partial \H{j}\over \partial u_n})$
Using the set of the  definitions above, one can easily prove the following
\hfill
\leftline{\bf Proposition}
\smallskip
Operator $\cal R$ defines the hierarchy of compatible Poissson structures
$${\Omega }^{(j)}={\cal R}^{2j-1}\cdot {\Omega }^{({1\over 2})}\eqno(A.4)$$
such that
$$\{\H{i},\pmatrix{p_k\cr u_k\cr}\}_j=
\{\H{i'},\pmatrix{p_k\cr u_k\cr}\}_{j'}
\qquad\hbox{when}\qquad i+j=i'+j'$$
\smallskip
$${\{\H{i},\H{j}\}}_l=0\qquad \forall \; i,\;j,\;l$$

\vfill\eject

\centerline{\bf References}

\bigskip

{\obeylines\smallskip
1. L. Faddeev, L. Takhtadjan, Lect. Notes in Phys. {\bf 246} (1986), 166

\smallskip

2. O. Babelon, Phys.Lett. {\bf B238} (1990), 234;

\smallskip

\qquad O. Babelon, L. Bonora, Phys.Lett. {\bf B253} (1991), 365

\smallskip

3. A. Volkov, Zap.Nauch.Sem.LOMI {\bf 150} (1986), 17; ibid. {\bf 151} (1987)
24;
\qquad Theor. Math. Phys. {\bf 74} (1988) 135

\smallskip

4. F. Falceto, K. Gawedzki, {\it Lattice Wess-Zumino-Witten models and Quantum
Groups},
\qquad preprint {\bf IHES/P/92/73} (1992)

\smallskip

5. A. Zamolodchikov, JETP Lett. {\bf 46} (1987), 160; Adv.St.Pure Math.
{\bf 19} (1989), 641;
\qquad Rev.Math.Phys. {\bf 1} (1990), 197

\smallskip

6. O. Babelon, Phys.Lett. {\bf B215} (1988), 523

\smallskip

7. H. Braden et al., Nucl.Phys. {\bf B338} (1990), 689

\smallskip

8. A. Belov, K. Chaltikian, "{\it Lattice analogues of W-algebras and Classical
Integrable Equations}",

\qquad preprint {\bf ILG-TMP-93-01}, hep-th/9303166

\smallskip

9. B. Feigin, Talk at September Sankt-Peterburg meeting on Geometry and
Physics,
 1992
\smallskip

10. V. Fateev, S. Lukyanov, Zh.E.T.P. {\bf 94}(3) (1988), 23

\smallskip

\qquad A. Gerasimov, {\it "Quantum Group Structure in Minimal Models"},
preprint
 {\bf ITEP-91-22}
(1991);

\smallskip

\qquad C. Gom\'ez, G. Sierra, {\it "Quantum Group Symmetry of Rational
Conformal
 Field Theories"},

\qquad  preprint {\bf UGVA-DPT-90/04-669} (1990)

\smallskip

11. K. Ueno, K. Takasaki, Adv. Studies in Pure Math. {\bf 4} (1984), 1

\smallskip

12. L. Faddeev, L. Takhtadjan,"{\it Hamiltonian Aprroach in Soliton Theory}"
(Springer, Berlin 1987
)

\smallskip

13. C. Xiong, Phys. Lett. {\ bf B279} (1992), 347

\smallskip

14. L. Bonora, C. Xiong, "{\it Multi-matrix models without continuous limit}",

\qquad preprint {\bf SISSA-ISAS 211/92/EP}, (1992)

\smallskip

15. F. Yu, Y.-S. Wu, Phys. Rev. Lett. {\bf 68} (1992), 2996

\smallskip}

\end